\documentclass[aps,prl,twocolumn,showpacs,groupedaddress]{revtex4} 
\usepackage{graphicx}  
\usepackage{dcolumn}   
\usepackage{bm}        
\usepackage{amssymb}   
\usepackage{amsmath}

\begin{document}

\title{Unconventional Transport in the `Hole'  Regime of a Si Double Quantum Dot}
\author{Teck Seng Koh, C. B. Simmons, M. A. Eriksson, S. N. Coppersmith, Mark Friesen}
\affiliation{Department of Physics, University of Wisconsin-Madison, Madison, WI 53706, USA}
\date{Mar 12, 2011.}

\begin{abstract}
Studies of electronic charge transport through semiconductor double quantum dots rely on a conventional `hole' model of transport in the three-electron regime. We show that experimental measurements of charge transport through a Si double quantum dot in this regime cannot be fully explained using the conventional picture. {Using a Hartree-Fock (HF) formalism and relevant HF energy parameters extracted from transport data in the multiple-electron regime,} we identify a novel spin-flip cotunneling process that lifts a singlet blockade.
\end{abstract}

\maketitle 

\begin{figure*}[!th]
\includegraphics[scale=0.22]{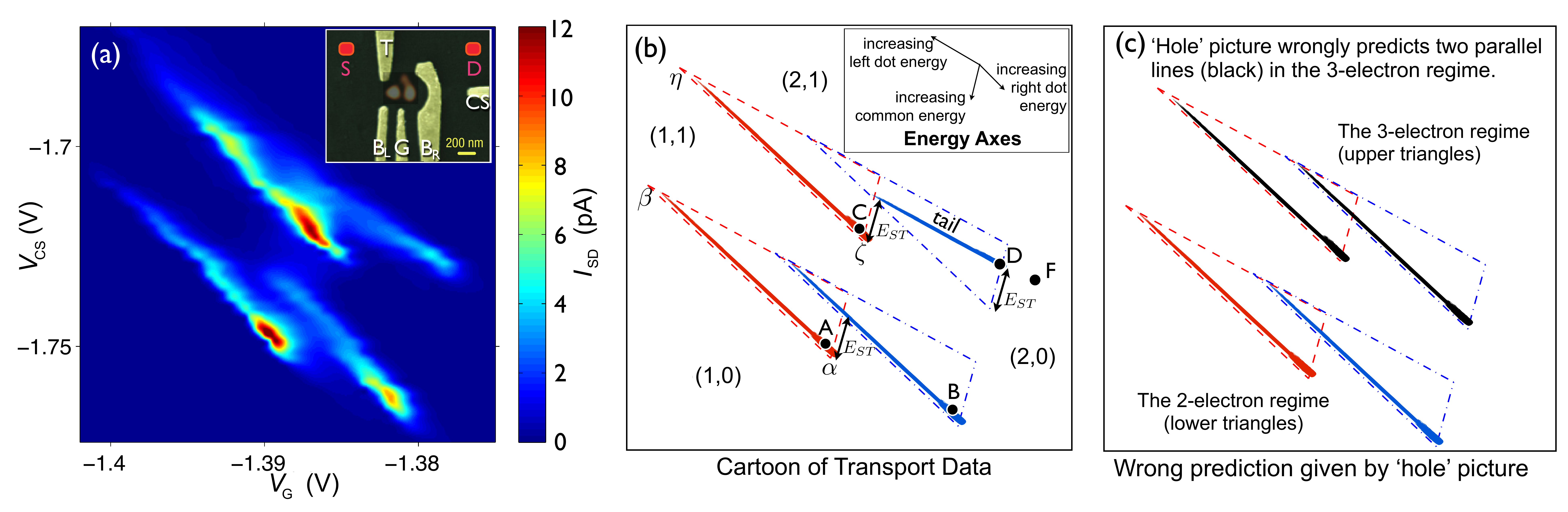}
\caption{\label{fig:data}
(a) Transport current $I_{\text{SD}}$ in a Si/SiGe double quantum dot (color scale) as a function of controlling gate voltages, $V_{\text{G}}$ (V) and $V_{\text{CS}}$ (V), reported in Ref.~\onlinecite{Nakul2008}. Inset shows an SEM image of the gates with a numerically simulated double dot overlayed. White letters (T, $\text{B}_{\text{L}}$, G, $\text{B}_{\text{R}}$ and CS) label gates and red letters (S, D) label the source and drain. 
For this data, electrons flow from left to right.
(b) A cartoon of the bias triangles and lines of high current. Inset shows the energy axes of the dots.  The lower (upper) features are the electron (hole) triangles. Red dashed lines represent current through the singlet (singlet-like) channel of the electron (hole) bias triangle and blue dot-dashed lines the triplet (triplet-like) triangle. A and B are resonant peaks of the singlet and triplet electron triangles. C is the resonant peak of the singlet-like hole triangle. A and C are the triple points at the boundary of the (1,0), (1,1), (2,0) and (1,1), (2,0), (2,1) charge occupations. 
F lies along the line extending from the tail; it is a representative point where cotunneling is dominant.
$E_{ST}$ is the (2,0) singlet-triplet energy splitting. Data are obtained at a reverse-bias source-drain voltage, $V_{\text{SD}}= -0.274$ mV, first published in Ref.~\onlinecite{Nakul2008} as $-0.3$ mV. Ref.~\onlinecite{Christie2010} details the quantitative fits to identify the triangles. 
(c) The prediction using the conventional hole picture in the three-electron regime is shown as two parallel lines (black), which is inconsistent with the tail observed in the data.
}
\end{figure*}

In quantum computing, semiconductor quantum dots have long been considered as good candidates for qubits~\cite{QD1,QD2,QD3}. A promising architecture for such qubits is the double quantum dot~\cite{QD3,DQD1,DQD2}. Understanding spin-dependent transport~\cite{EnDep1,EnDep2,Nakul2008,SB1,SB2} is important for using the spin degree of freedom in a double dot qubit. Here, we show that transport data taken in the three-electron regime of a double dot in a Si/SiGe heterostructure have features that are qualitatively inconsistent with the conventional model of `hole' transport~\cite{vanderWiel}, because this model does not account for transport through excited states. {Using the Hartree-Fock (HF) formalism with singly excited configurations~\cite{Szabo}, together with relevant HF parameters extracted from the transport data (see \cite{Christie2010}),  we 
demonstrate that} the striking features in the data arise from a novel spin-flip cotunneling process in which the multi-electron nature of the system enters fundamentally.

Several experiments have probed charge transport through double quantum dots in the few-electron regime
and investigated effects such as energy-dependent tunneling and spin-dependent transport~\cite{EnDep1,EnDep2,Nakul2008,SB1,SB2}. Transport in the three-electron regime is well-described in terms of holes when all the intra-dot relaxation rates are much faster than the interdot tunnel rate, so that the dominant transport channels are through the lowest energy states of each dot, as is typically the case in GaAs devices~\cite{SB1,vanderWiel}.

Our theoretical work is based on data~\cite{Nakul2008}, in which a lateral double quantum dot was formed by electrostatic gating of a Si/SiGe heterostructure, as shown in the inset of Fig.~\ref{fig:data}(a). Fig.~\ref{fig:data}(a) shows source-drain current versus controlling gate voltages at a fixed source-drain bias voltage. Transport through the two dots is energetically favorable within triangular regions whose size is determined by the source-drain bias. Lines of high current in these bias triangles are associated with fast tunneling between the dots and between the dots and the leads~\cite{vanderWiel}. From the orientation of the line $\alpha\beta$ in Fig.~\ref{fig:data}(b), we deduce that it is associated with the resonance of an energy level in the `left' dot with the chemical potential in the left lead. Quantitative fits allow the edges of the triangles to be determined and are reported in detail in Ref.~\onlinecite{Christie2010}. Fig.~\ref{fig:data}(b) is a schematic diagram of the bias triangles, with energy axes shown in the inset. The lower features arise from transport when the double dot contains either one or two-electrons (`two-electron' regime), while the upper features reflect transport when the dot contains either two or three-electrons (`three-electron' regime, also conventionally termed `hole' regime).

There are two regions of current flow in each transport regime, shown in Fig.~\ref{fig:data}(a).  Each of these regions of current is contained in a triangle, shown in either blue or red in Fig.~\ref{fig:data}(b).  The presence of current in the blue triangle implies that there is significant transport through excited states of the dots~\cite{Nakul2008, Christie2010}, something that has recently been observed in transport through a single phosphorous donor in silicon as well~\cite{LansbergenPreprint}.  

Because the effective electronic mass in Si is much larger than in GaAs, transport is energetically favorable within each bias triangle, but the triangle is not entirely filled because the electron tunneling rate is strongly energy-dependent~\cite{EnDep1,EnDep2,Nakul2008}. The two parallel lines of high current that are observed along the left edges of the singlet and triplet triangles in the two-electron regime (lower feature) indicate that energy-dependent tunneling across the left barrier is the bottleneck in the total tunneling rate~\cite{Christie2010}. 

In the three-electron regime, the conventional picture that describes the conduction in terms of holes predicts that there should be two parallel lines of high current (Fig.~\ref{fig:data}(c)). This is because, in the two-electron regime, electron occupancy cycles through the states (1,0) $\rightarrow$ (2,0) $\rightarrow$ (1,1), and the three-electron regime is modeled conventionally~\cite{vanderWiel} as hole transport in the opposite direction: (1,1) $\rightarrow$ (0,2) $\rightarrow$ (0,1), where the numbers represent electron or hole occupancy in the left and right dots. Due to particle-hole symmetry, the hole picture predicts parallel lines of high current similar to the data in the two-electron regime (Fig.~\ref{fig:data}(c)).

The transport data in Fig.~\ref{fig:data}(a) is inconsistent with a picture in terms of holes, as it shows two lines of high current in the three-electron regime (upper feature) that are clearly not parallel. In this regime, there is a line of high current at the left edge of the bias triangle (line $\zeta\eta$) for ground state transport, which is expected since the left barrier is observed to be the bottleneck in the two-electron regime. However, in the bias triangle for excited state transport, there is a `tail' parallel to the right edge but away from it, which the hole picture completely fails to describe. 

\begin{figure}[!t]
\includegraphics[scale=0.23]{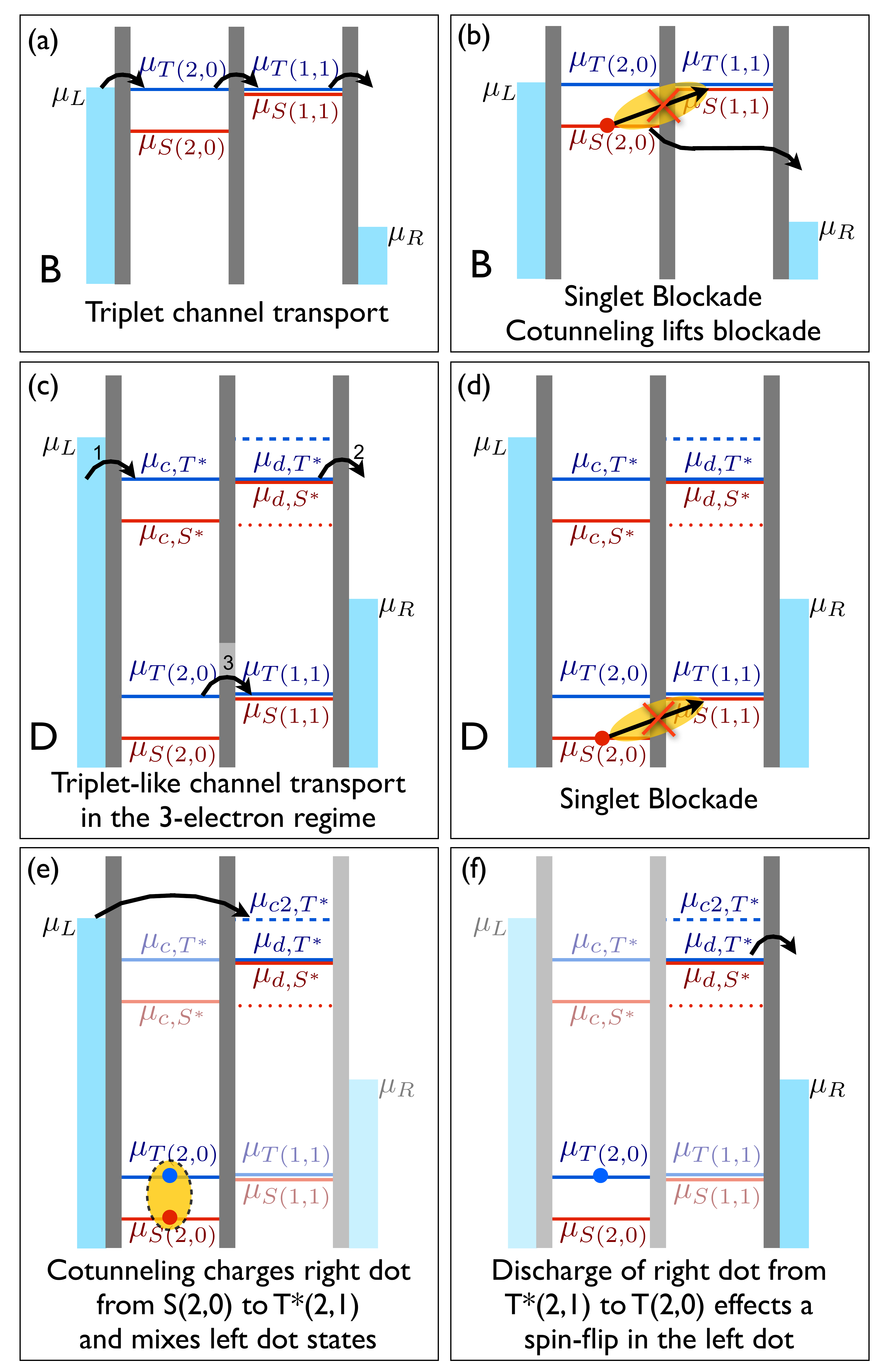}
\caption{\label{fig:levels} Diagrams for transport through excited states and the process of `spin-flip cotunneling.'
(a) Triplet channel transport in the two-electron regime. At B, the resonant peak of the triplet channel, transport is allowed through the triplet levels, $\mu_{T(2,0)}$ and $\mu_{T(1,1)}$.
(b) When the singlet level $\mu_{S(2,0)}$ is loaded, transport is energetically uphill and blockaded. Cotunneling of a left dot electron out 
to the right lead lifts the blockade to resume triplet channel transport.
(c) In the three-electron regime, transport occurs in the following cycle: 
step 1, $(1,1)\rightarrow(2,1)$; step 2,  $(2,1)\rightarrow(2,0)$; step 3, $(2,0)\rightarrow(1,1)$. At D, these occur through the triplet-like channel (blue).
(d) Because of the loading of the singlet-like $\mu_{c,S^*}$ level (red) at D, the system ends up in the $S(2,0)$ state, whereby transport is blockaded. 
(e) When an electron cotunnels from the left lead into the right dot to form a triplet-like state, it puts the left dot into an admixture of singlet and triplet. 
(f) The right dot discharges from $\mu_{d,T^*}$, leaving a triplet state in the left dot, thus causing a spin-flip and resuming transport. 
}
\end{figure}

To understand the problem theoretically, we formulate it in terms of chemical potentials and use the Hartree-Fock (HF) approximation with singly excited configurations to determine the spin eigenfunctions and energy levels of each of the double dot states involved in the three-electron regime. 
{The relevant parameters in the HF formulation are extracted from the transport data as detailed in Ref.~\onlinecite{Christie2010}.}
From the energy levels and possible transitions between states, we calculate the electrochemical potentials for charging or discharging a dot by one electron~\cite{Supp}. The four relevant electrochemical potentials for the dots are shown for the two-electron case in Fig.~\ref{fig:levels}(a,b). For the three-electron case, the full set of ten electrochemical potentials, shown in Fig.~\ref{fig:levels}(c,d), is clearly greater than the four electrochemical potentials for transport 
modeled on two holes. The many-electron nature of the problem thus enters our analysis of transport naturally.

Without going into the details of the HF calculations, we can gain some insight into the possible (2,1) states using qualitative arguments. The pure singlet 
and triplet states, $S(2,0)$ and $T(2,0)$,
are no longer orthogonal when we include a weak coupling to a third electron in the right dot. The perturbation 
leads to a `singlet-like' ground state $S^*(2,1)$, whose spin configuration in the left dot is mainly $S(2,0)$ with a small admixture of $T(2,0)$.
The $S^*(2,1)$ state has spin $S_z=\pm 1/2$ and is doubly degenerate.  
The perturbation also leads to `triplet-like' states $T^*(2,1)$, for which spin addition gives $S_z=\pm1/2$ or $\pm3/2$. The $S_z=\pm1/2$ states contain mainly triplet $T(2,0)$ with a small admixture of $S(2,0)$.
The $S_z=\pm3/2$ states have spins that are either all up or all down;  
they are doubly degenerate without any admixture of singlet states. 
The triplet degeneracies are lifted due to the fact that exchange energies are different for different three-electron spin configurations. 
The energy splittings arise from inter-dot interactions, which are much smaller than intra-dot interactions.
Thus, the splittings within the triplet-like manifold are much finer than the splitting between the singlet- and triplet-like manifolds. These arguments are borne out by our calculations~\cite{Supp}. 

From the energy levels calculated with the HF Hamiltonian, we can explain how the electrochemical potentials, shown in Fig.~\ref{fig:levels}(c), are obtained. In the three-electron regime, electron occupancy 
cycles through $(1,1)\rightarrow(2,1)\rightarrow(2,0)$. The first transition corresponds to charging of the left dot from a (1,1) to a (2,1) state. For clarity, we do not distinguish between the two closely spaced (1,1) energies, nor do we distinguish between the three closely spaced $T^*(2,1)$ energies. We therefore obtain two distinct electrochemical potentials, $\mu_{c,T^*}$ and $\mu_{c,S^*}$, shown in Fig.~\ref{fig:levels}(c), which are the energies needed to charge the left dot from a (1,1) state to the $T^*(2,1)$ and $S^*(2,1)$ states respectively. The second transition represents the discharge of an electron from a (2,1) to a (2,0) state. Electrochemical potentials, $\mu_{d,T^*}$ and $\mu_{d,S^*}$, drawn on the right dot, represent the discharge of the right dot from $T^*(2,1)$ to $T(2,0)$ and $S^*(2,1)$ to $S(2,0)$ respectively. These are the continuous lines (blue and red) on the right dot in Fig.~\ref{fig:levels}(c). Due to singlet-triplet mixing in the left dot, two other transitions of much smaller likelihood are possible. They are the $S^*(2,1)$ to $T(2,0)$ and $T^*(2,1)$ to $S(2,0)$ transitions, represented by the red dotted and blue dashed levels respectively, in the same figures. The last step in the cycle is the inter-dot transition, $(2,0)\rightarrow(1,1)$. The chemical potentials in this step are identical to the two-electron case (Fig.~\ref{fig:levels}(a,b)) and are labeled as $\mu_{S, T(2,0)}$ and $\mu_{S,T (1,1)}$.

We can now explain the tail in the transport data, 
which, as described above, is a prominent feature that is qualitatively inconsistent with a description in terms of holes. At point D in Fig.~\ref{fig:levels}(c), transport is allowed through the blue triplet-like levels. However, it is also possible to load the red singlet-like $\mu_{c,S^*}$ level.
In this case, as the right dot discharges, the system is likely to end up in the $S(2,0)$ state, 
where transport is energetically uphill and therefore blockaded (Fig.~\ref{fig:levels}(d)). We call this a `singlet blockade.' The lifting of the singlet blockade along the tail is shown in sequence in Figs.~\ref{fig:levels}(e) and (f). Starting from $S(2,0)$, the double dot forms a triplet-like $T^*(2,1)$ state when an 
electron from the left lead cotunnels into the right dot, as shown in Fig.~\ref{fig:levels}(e). The charging of the right dot in this transition requires the same energy as its reverse discharging process ($T^*(2,1)$ to $S(2,0)$), 
represented by the blue dashed line in Fig.~\ref{fig:levels}(c). It is
labeled by $\mu_{c2,T^*}$ in Figs.~\ref{fig:levels}(e) and (f). Because the triplet-like 
state contains an admixture of singlet and triplet states in the left dot, when the right dot discharges from the $\mu_{d,T^*}$ level, the left dot ends up in the triplet (2,0) state, thus causing a spin-flip. With the singlet blockade lifted, the system then completes the cycle into the triplet (1,1) state and transport resumes as shown in Fig.~\ref{fig:levels}(c). We term this process `spin-flip cotunneling.'

The tail in the transport data in Fig.~\ref{fig:data}(a) is bright along its entire length because the chemical potentials for the right dot and the left lead are the same. Point D is the brightest point along the tail because of the fast inter-dot tunneling when $\mu_{T(2,0)}$ is aligned with $\mu_{T(1,1)}$. 

The spacing of the tail away from the edge of the triangle is consistent with the energy difference between the $\mu_{c2,T^*}$ and $\mu_{d,T^*}$ levels on the right dot (Fig.~\ref{fig:levels}(e)) being equal to $E_{ST}$, the (2,0) singlet-triplet energy splitting. To understand how this is consistent with the transport data, we start from point C in Fig.~\ref{fig:data}(b) and note that when both dot energies fall by $E_{ST}$, the blue, dashed $\mu_{c2,T^*}$ level of the right dot lines up with the Fermi level of the left lead. This measure of $E_{ST}$ is also consistent with other measures of ST splitting~\cite{Christie2010}.

The significant role cotunneling plays in the triplet and triplet-like transport channels of the two and three-electron regime is interesting. In both cases, cotunneling by itself does not contribute significantly to the current, but plays the role of allowing transport to resume by lifting the singlet blockade.

Current will flow through the triplet-like channel when the loading rate is comparable to the unloading rate in the singlet-like channel~\cite{Supp}. In the Supplementary Information, we estimate these rates and find that they are indeed the same order of magnitude. The blockade is therefore lifted about as quickly as it is encountered.
In this way, spin-flip cotunneling enables transport through the triplet-like channel.
The resulting current is that of the unblockaded, triplet-like channel, reduced by a factor of $\sim 2$~\cite{Supp}.

Interestingly, transport via the triplet channel was not observed in the experiments reported in Ref.~\onlinecite{SB1}. In that study, a conventional hole model was sufficient to describe transport in the three-electron regime, as consistent with the fact that transport occurred through the ground states in the two-electron regime.

It is also interesting to compare the intra-dot spin-flip times with inter-dot tunneling times for GaAs and Si. In GaAs devices, spin-flip times range from $\sim$200 $\mu$s for a two-electron dot~\cite{Fujisawa2002}, to $\sim$0.85 ms (at 8~T~\cite{Elzerman2004}) and $>1$~s (at 1~T and 120~mK~\cite{Amasha2008}) for single electron dots. Recent experiments report spin-flip times in single electron dots in Si~\cite{Xiao2010, Morello2010, Hayes2009, Simmons2010} ranging from 40~ms (at 2~T~\cite{Xiao2010}) to 6~s (at 1~T~\cite{Morello2010}), at low temperatures. The tunnel coupling for the same Si double dot studied here was found to be 10 ns (25 ns) in the elastic (inelastic) tunneling regime~\cite{Christie2010}. However, tunnel couplings for electrostatically gated semiconductor double dots are tunable and can be both larger or smaller than spin-flip times. 

{We note that a pulsed gate experiment~\cite{EnDep2} exhibiting phenomena arising from singlet-triplet mixing, as described above, is presented in \cite{Supp}.}

In summary, we have shown that the conventional hole model of transport in the three-electron regime fails qualitatively because of the importance of excited state transport. {The Hartree-Fock formalism, with relevant parameters fitted to transport data, leads to the description of a model which explains all of the features of the transport data, including a novel process of spin-flip cotunneling.}

We thank Nakul Shaji for useful discussions and gratefully acknowledge funding from ARO and LPS
(W911NF-08-1-0482) and NSF (DMR-0805045).

\appendix
\section{Supplementary Material: Unconventional Transport in the `Hole'  Regime of a Si Double Quantum Dot}

The following Supplementary Information presents our detailed calculations of the Hartree-Fock approximation and electrochemical potentials relevant for reverse bias transport in the three electron regime. We estimate the spin-flip cotunneling occurrence rate, quantify the effect spin-flip cotunneling has on transport current in the triplet-like triangle, and show that it is consistent with the data. We propose an experiment to verify the singlet-triplet mixing mechanism and also present a detailed analysis showing that at forward bias, transport data has a feature associated with the process of spin-flip cotunneling, as discussed in the main paper.

\section{Hartree-Fock Approximation with Singly Excited Configurations} 

The Hartree-Fock (HF) Hamiltonian used in our calculations is given by~\cite{Szabo}
\begin{equation} \tag{S1}
H=\sum_{i,j}{\langle i|h|j\rangle a_{i}^\dagger a_{j}}+\frac{1}{2}\sum_{i,j,k,l}{\langle ij| kl\rangle a_{i}^\dagger a_{j}^\dagger a_{l} a_{k}},
\end{equation}
where the sums run over the spin orbitals chosen to form the basis of the problem. Operators $a_{i}^\dagger$ and $a_{i}$ create and annihilate an electron in the $i$-th spin-orbital respectively. The first term contains the one electron operator, $h \equiv \sum -1/2\nabla^2_\alpha + V(\vec{x}_\alpha - \vec{x}_{dot})$, the sum of the kinetic and potential energies of each $\alpha$-th non-interacting electron in a quantum dot centered at $\vec{x}_{dot}$. We note that the parabolic well approximation is a reasonable form of the potential for our geometry. The second term contains the two electron Coulomb operator, $ \langle ij | kl \rangle \equiv \int \, d\bold{x}_\alpha d\bold{x}_\beta \chi_i^*(\bold{x}_\alpha)\chi_j^*(\bold{x}_\beta) r^{-1}_{\alpha\beta} \chi_k(\bold{x}_\alpha)\chi_l(\bold{x}_\beta) $, where $\chi_{i}(\bold{x_\alpha})$ denotes the $i$-th spin-orbital wave-function of the $\alpha$-th electron, following the notation of Ref.~\onlinecite{Szabo}.

We define the quantities $s_{ij} \equiv \langle i|j \rangle$, the overlap between the $i$-th and $j$-th spin orbitals; $e_i  \equiv  \langle i|h|i \rangle$, the single particle energy; $J_{ij} \equiv \langle ij | ij \rangle$ and $K_{ij}  \equiv \langle ii | jj \rangle$, the Coulomb and exchange interactions respectively. $ t_{ij} \equiv \langle i|h|j \rangle$ and $ \Gamma^{ij}_{lk}\equiv \langle ij |  kl \rangle$ govern coherent tunneling between dots and are first and second order in the overlap integral, respectively. 

Singly-excited configurations that form the basis for the (2,1) Hamiltonian matrix, are constructed from dot-centered single particle spatial orbitals $\{1, 2\}$ and $\{3, 4\}$, centered on the left and right dots respectively. The two sets of spatial orbitals are not orthogonal to each other, due to the overlap across the finite barrier between the dots. Within each set, however, the spatial orbitals are orthonormal. The corresponding spin orbitals are denoted by $\{1, \bar{1}, 2, \bar{2}\}$ and $\{3, \bar{3}, 4, \bar{4}\}$ where the overbar (or lack of) denote spin down (up). Table~\ref{tab:basis} shows the basis states used to construct the $10\times10$ Hamiltonian matrix.

\begin{table}
\caption{\label{tab:basis}
Basis states used for the (2,1) configuration.}
\begin{ruledtabular}
\begin{tabular}{llc}
\multicolumn{2}{c}{Basis states}&Spin \\
in spin-orbital notation & in $S,T$ notation& $S_z$\\
\hline
$|123\rangle$  & $|T_+(2,0)\rangle\otimes|3\rangle$ &3/2\\
$|12\bar{3}\rangle$  & $|T_+(2,0)\rangle\otimes|\bar{3}\rangle$ &1/2\\
$(|\bar{1}23\rangle+|1\bar{2}3\rangle)/\sqrt{2}$  & $|T_0(2,0)\rangle\otimes|3\rangle$ &1/2\\
$(|\bar{1}23\rangle-|1\bar{2}3\rangle)/\sqrt{2}$  & $|S_1(2,0)\rangle\otimes|3\rangle$ &1/2\\
$|1\bar{1}3\rangle$  & $|S_0(2,0)\rangle\otimes|3\rangle$ &1/2\\
\multicolumn{2}{c}{and their spin reversed counterparts}&\\
\end{tabular}
\end{ruledtabular}
\end{table}

The (2,1) Hamiltonian consists of four block diagonals: two $1\times1$ blocks corresponding to a total spin of $S_z=\pm3/2$, and two $4\times4$ blocks corresponding to $S_z=\pm1/2$. In the $S_z=\pm1/2$ blocks, off-diagonal matrix elements are second order in the overlap between spin orbitals centered on different dots and represent the amount of mixing of the singlet and triplet states in the left dot. As such, we term the ground state, `singlet-like', and denote it as $S^*(2,1)$. It contains mainly singlet $S(2,0)$ with some admixture of triplet $T(2,0)$ in the left dot, coupled to the third electron in the right dot. The first excited state is the pure $S_z=\pm3/2$ state with all spins either up or down, which does not have any singlet-triplet mixing. The next two higher states, which are closely spaced with each other as well as with the first excited state, contain a small admixture of singlet with mainly triplet states in the left dot, coupled to the right dot. For brevity, we term the three excited states triplet-like $T^*(2,1)$ states, and do not distinguish between them henceforth, as the energy separation between the triplet-like states is of second order in the overlap integral and is small compared to the spacing between the ground and first excited state. The next higher state is well separated from these and does not enter into our analysis.

The $S_z=1/2$ block of the Hartree-Fock Hamiltonian for the (2,1) state in the basis $ |S_0(2,0)\rangle\otimes|3\rangle, |T_0(2,0)\rangle\otimes|3\rangle, |S_1(2,0)\rangle\otimes|3\rangle, |T_+(2,0)\rangle\otimes|\bar{3}\rangle $ is
\begin{widetext}
\begin{equation}\tag{S2}
\begin{scriptsize}
\hat{H}_{(2,1)}^{S_z=1/2}=
\left(\!
\begin{array}{cccc}
 2 e_1+e_3+J_{11}+2 J_{13} &(\Gamma _{23}^{13}+s_{13} t_{23}+s_{23}t_{13})/\sqrt{2} &(2 \Gamma_{12}^{11}-\Gamma _{23}^{13}+2 \Gamma _{32}^{13} & -\Gamma _{23}^{13}-s_{13}t_{23}-s_{23}t_{13} \\
 -K_{13}+2 s_{13} t_{13} &  & -s_{13} t_{23} - s_{23} t_{13})/\sqrt{2} & \\
 & & &\\
 
(\Gamma _{23}^{13}+s_{13} t_{23}+s_{23}t_{13})/\sqrt{2} & e_1+e_2+e_3+J_{12}+J_{13}+J_{23}& (K_{23}-K_{13})/2 & -(K_{13}+K_{23}\\
  & -K_{12}-(K_{13}+K_{23})/2  & - s_{13} t_{13}+ s_{23} t_{23} & +2 s_{13} t_{13} +2s_{23}t_{23})/\sqrt{2} \\
  & -s_{13} t_{13}-s_{23} t_{23} & &\\
  & & &\\

(2 \Gamma _{32}^{13}-\Gamma _{23}^{13}+2 \Gamma _{12}^{11} &(K_{23}-K_{13})/2 & e_1+e_2+e_3+J_{12}+J_{13}+J_{23}&
   (K_{13}-K_{23}\\
 -s_{13} t_{23} - s_{23} t_{13})/\sqrt{2}  & -s_{13} t_{13}+ s_{23} t_{23} & +K_{12}-(K_{13}+K_{23})/2 & +2 s_{13} t_{13} -2s_{23}t_{13})/\sqrt{2}  \\
   & & -s_{13} t_{13}-s_{23} t_{23} &\\
   & & &\\

 -\Gamma _{23}^{13}-s_{13}t_{23}-s_{23}t_{13} & -(K_{13}+K_{23} &(K_{13}-K_{23}
   & e_1+e_2+e_3+J_{12}\\
 & +2 s_{13} t_{13} +2s_{23}t_{23})/\sqrt{2}&+2 s_{13} t_{13} -2s_{23}t_{13})/\sqrt{2} &+J_{13}+J_{23}-K_{12} 
\end{array}
\! \right)
\end{scriptsize}
\end{equation}
\end{widetext}

\begin{figure}
\includegraphics[scale=0.25]{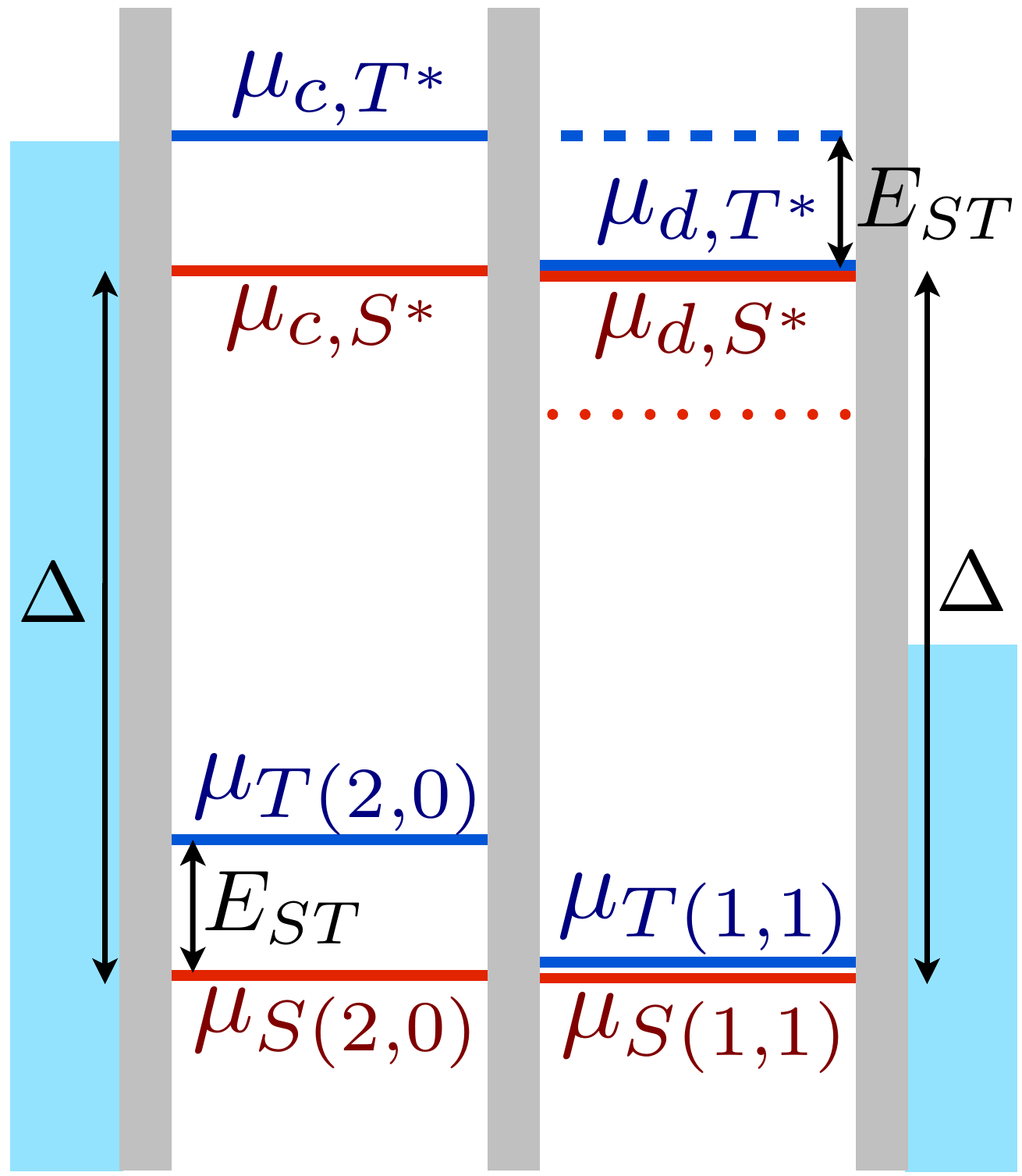}
\caption{\label{fig:echem} 
Electrochemical potentials $\mu$ for different states relevant to transport in the three electron regime. Here, the label $E_{ST}$ is the (2,0) singlet-triplet splitting. $\Delta$ is the spacing between the chemical potentials of the two and three electron regimes, as defined in Eq.~\ref{eq:delta}, and is equal to the energy spacing between points A and C in the transport data in Fig.~1(b) of the main text. Both quantities are extracted from the transport data as detailed in Ref.~\onlinecite{Christie2010}, giving $E_{ST}\approx174$~$\mu$eV for (2,0) singlet-triplet splitting and $\Delta\approx~500~\mu$eV. 
}
\end{figure}

The off-diagonal elements are second order in the overlap integral and therefore small, so the problem can be treated perturbatively. The energies, to first order approximation, are given in Table~\ref{tab:energies}. The splitting between the singlet-like and triplet-like states is approximately equal to the singlet-triplet splitting of the (2,0) state, $E_{ST(2,0)}$. We apply the HF Hamiltonian (Eq.~S1) to calculate the singlet and triplet energies of the two electron states, which are necessary for the calculation of electrochemical potentials. Similar analyses using molecular and dot-centered orbitals have been done elsewhere~\cite{HF2}, so we do not reproduce the calculations here. We refer the reader to Section III.~D of Ref.~\onlinecite{HF2} for calculations using dot-centered orbitals. We caution the reader that, in this previous study, the double dot geometry and material are different, so their numerical results will not be applicable for our case. However, the expressions for energy levels in terms of dot-centered single-particle, direct and exchange Coulomb energies in Eq.~(18--21) of Ref.~\onlinecite{HF2} are relevant for our two electron double dot case.


\begin{table}
\caption{\label{tab:energies}
Energies and characteristics of the (2,1) states.}
\begin{ruledtabular}
\begin{tabular}{llcc}
Energy (ascending order)& State & Spin $S_z$ \\
\hline
$2e_1+e_3+J_{11}+2J_{13}$ & Singlet-like,&$\pm1/2$ \\
$-K_{13}+2s_{13}t_{13}$& Mixed&\\
 &&\\
$e_1+e_2+e_3+J_{12}+J_{13}$ & Triplet-like,&$\pm$3/2 \\
  $+J_{23}-(K_{12}+K_{13}+K_{23})$ & Pure&\\
 &&\\
$e_1+e_2+e_3+J_{12}+J_{13}+J_{23}-K_{12}$ & Triplet-like,&$\pm$1/2 \\
$-(K_{13}+K_{23})/2-(s_{13}t_{13}+s_{23}t_{23})$& Mixed&\\
&&\\
$e_1+e_2+e_3+J_{12}+J_{13}+J_{23}-K_{12}$ & Triplet-like,&$\pm$1/2 \\
& Mixed &\\
\end{tabular}
\end{ruledtabular}
\end{table}

Once the expressions for the energies are known, it is straightforward to calculate the relevant expressions for electrochemical potentials. For clarity, we neglect the singlet-triplet splitting in the (1,1) states and the energy spacing between the three triplet-like (2,1) states and obtain two distinct electrochemical potentials. The electrochemical potentials for charging the left dot from a (1,1) state to the $S^*(2,1)$ and $T^*(2,1)$ states are, respectively, 
\begin{align}
\mu_{c,S^*} &= e_1 + J_{11} + J_{13} - E_{ST(1,1)} + E_{Ldot},\tag{S3}\\
\mu_{c,T^*} &\approx \mu_{c,S^*} + E_{ST(2,0)}.\tag{S4}
\end{align}
$E_{Rdot}$ ($E_{Ldot}$) is the electrostatic energy of the right (left) quantum dot. The electrochemical potentials for discharging the right dot from the triplet-like $T^*(2,1)$ to the triplet $T(2,0)$ state and from the singlet-like $S^*(2,1)$ to singlet $S(2,0)$ state are
\begin{equation}\tag{S5}
\mu_{d,T^*} \approx \mu_{d,S^*} = e_3 +2J_{13}-K_{13}+E_{Rdot}.
\end{equation}

There are two more electrochemical potentials involved in the discharge of the right dot from the (2,1) states, due to the mixing of the singlet and triplet (2,0) states. These are associated with the discharge from $T^*(2,1)$ to $S(2,0)$ and $S^*(2,1)$ to $T(2,0)$, and are shown by the blue dashed and red dotted lines on the right dot in Fig.~\ref{fig:echem}. They are separated from $\mu_{d,T^*}$ and $\mu_{d,S^*}$ by the (2,0) singlet-triplet splitting $E_{ST}$.

The last set of electrochemical potentials needed are those for the two electron regime, which are given by
\begin{align}
\mu_{S(2,0)} &= e_1 +J_{11} + E_{Ldot},\tag{S6}\\
\mu_{T(2,0)} &= \mu_{S(2,0)} + E_{ST(2,0)},\tag{S7}\\
\mu_{T(1,1)} &= e_3 + J_{13} - K_{13} + E_{Rdot},\tag{S8}\\
\mu_{S(1,1)} &= \mu_{T(1,1)} - E_{ST(1,1)}.\tag{S9}
\end{align}

The chemical potentials for the two and three electron regimes satisfy the relation
\begin{equation}
\label{eq:delta0}\tag{S10}
\mu_{c,S^*} - \mu_{S(2,0)} \approx \mu_{d,S^*} - \mu_{S(1,1)},
\end{equation}
and we therefore define 
\begin{equation}
\label{eq:delta}\tag{S11}
\Delta \equiv \mu_{d,S^*} - \mu_{S(1,1)}.
\end{equation}

Here, we point out that it is the relative positions of the electrochemical potentials that are important in order to build the transport model shown in Fig.~\ref{fig:echem}. The important energy quantities are therefore, the (2,0) and (1,1) singlet-triplet energy splittings $E_{ST(2,0)} \approx 174 ~\mu$eV and $E_{ST(1,1)} \approx 4 ~\mu$eV, and the spacing $\Delta \approx 500 ~\mu$eV between the two and three electron triangles. These energies are extracted from the transport data as detailed in Ref.~\onlinecite{Christie2010}.

\section{Effect of Spin-Flip Cotunneling on Transport Current}

In this section, we show that current can flow through the triplet channel in the `hole' transport regime when certain conditions are met for the singlet channel.
Specifically, the singlet unloading rate should be equal to or larger than the singlet loading rate.  
We then show that these conditions are met in our data.

\subsection{Conditions on the singlet tunnel rates}
We first obtain the conditions on the singlet transport rates by developing rate equations, in analogy with those found in Ref.~\onlinecite{Nakul2008}.

We consider the contribution of loading and unloading rates of each channel to the total transport rate through the channel. Starting from the (1,1) state, the double dot is loaded into one of the (2,1) states from the left lead. The mean loading rate is given by
\begin{align}
\Gamma_{\text{load}} &\equiv 1/T_{\text{load}} \notag\\
                        &= \Gamma_{S\text{ load}} +\Gamma_{T\text{ load}} ,\tag{S17}
\end{align}
where $\Gamma_{S/T\text{ load}}$ are the loading rates into the singlet-like and triplet-like states from the left lead respectively.

The total time required for a complete transport cycle is
\begin{equation}\tag{S18}
T = T_{\text{load}} + p_T/ \Gamma_{T\text{ unload}} + p_S/ \Gamma_{S \text{ unload}} ,
\end{equation}
where  $\Gamma_{S/T\text{ unload}}$ are the unloading rates of the singlet-like and triplet-like channels respectively, and
\begin{align}
p_S &= \frac{\Gamma_{S\text{ load}}} {\Gamma_{S\text{ load}}+\Gamma_{T\text{ load}}},\tag{S19}\\
p_T &= \frac{\Gamma_{T\text{ load}}} {\Gamma_{S\text{ load}}+\Gamma_{T\text{ load}}},\tag{S20}
\end{align}
are the probabilities of loading the singlet-like and triplet-like channels.

In the triplet-like transport regime, the singlet-like channel is strongly suppressed by energy dependent tunneling.  If we assume that the loading and unloading rates for the triplet-like channel are much faster than the loading and unloading rates for the singlet-like channel, and if we also assume that the unloading rate of the triplet-like channel is much faster than its loading rate due to asymmetric tunnel barriers (see Ref.~\onlinecite{Nakul2008} for a discussion), then Eq.~S(18) reduces to
\begin{align}
T&\approx \left( 1+ \frac{\Gamma _{S \text{ load}}}{\Gamma _{S \text{ unload}}} \right)\frac{1}{\Gamma _{T \text{ load}}}+\frac{1}{\Gamma _{T \text{ unload}}}\notag\\
   &\approx \left( 1+ \frac{\Gamma _{S \text{ load}}}{\Gamma _{S \text{ unload}}} \right)\frac{1}{\Gamma _{T \text{ load}}}\tag{S21}.
\end{align}

In the limit of only one fast triplet-like channel, the total tunneling time is $T \approx 1/\Gamma _{T \text{ load}}$.  This implies that if $ \Gamma _{S \text{ load}} \ll \Gamma _{S \text{ unload}} $, then the current through the double dot is essentially a triplet-like channel current.  Therefore, it is important to determine the ratio of the loading to unloading rates of the singlet-like channel.

\subsection{Loading rates}
We now estimate the loading rates for the singlet-like and triplet-like channels at point D on the `tail'  of the transport data (Fig.~1, main text). Starting from the (1,1) charge configuration, the loading rates into the (2,1) states are strongly energy-dependent at the left barrier~\cite{Christie2010} and are given by 
\begin{equation}\tag{S22}
\Gamma_{S/T,\text{ load}} = \Gamma_0 e^{-E_{S/T} / E_L},
\end{equation}
where the amplitude $\Gamma_0 \approx 1.5\times10^8 \text{ s}^{-1}$, the energy dependent coefficient $E_L \approx 40~\mu\text{eV}$, $E_T\approx E_{ST(2,0)} \approx 174~\mu\text{eV}$~\cite{Christie2010} and $E_S \approx 2E_{ST(2,0)}$ at point D.

The loading rates of the triplet-like and singlet-like states at point D are therefore
\begin{align}
\Gamma_{T, \text{ load}} \approx 1.9\times10^{6} \text{ s}^{-1},\tag{S23}\\
\Gamma_{S, \text{ load}} \approx 2.5\times10^{4} \text{ s}^{-1}.\tag{S24}
\end{align}

\subsection{Unloading rates}
We now estimate the unloading rates for the singlet-like and triplet-like channels at point D.
Unloading of the triplet-like (2,1) state occurs when the electron in the right dot tunnels out across the right barrier, followed by interdot tunneling from the (2,0) to the (1,1) state. The tunneling rate across the right barrier $\Gamma_R \approx 9.2\times10^{9} \text{ s}^{-1}$ and the coherent interdot tunneling rate $\Gamma_i \approx 7.7\times10^{8}\text{ s}^{-1}$~\cite{Christie2010}. The unloading rate for the triplet-like channel is therefore given by
\begin{equation}\tag{S25}
\Gamma_{T,\text{ unload}} = (\Gamma_R^{-1} + \Gamma_i^{-1})^{-1}
                         \approx 7.1\times10^{8} \text{ s}^{-1}.
\end{equation}

The unloading rate of the singlet-like state is dominated by spin-flip cotunneling, which involves two low-probability processes:  the spin-flip and the cotunneling.  The spin-flip can be understood as the small probability $p_m$ that the $T^*$ state contains the $(2,0)$ singlet.  We then have
\begin{equation} \tag{S26}
\Gamma_{S,\text{ unload}} = p_m \Gamma_C , 
\end{equation}
where $\Gamma_C$ is the total cotunneling rate.

We can estimate $p_m$ from the (2,1) Hamiltonian of Eq.~(S2) by considering the subspace spanned by the states $|S_0(2,0)\rangle \otimes |3\rangle$ and $|T_0(2,0)\rangle \otimes |3\rangle$. The Hamiltonian within this subspace is given by the top-leftmost $2\times2$ matrix block in the right-hand side of Eq.~(S2). The off-diagonal element in this subspace is $(\Gamma_{23}^{13} +s_{13}t_{23} + s_{23}t_{13})/\sqrt{2} \approx \Gamma_{23}^{13}/\sqrt{2}$. The singlet-triplet mixing probability is therefore
\begin{equation}\tag{S27}
p_m \approx \left[ \frac{\Gamma^{13}_{23}} {2E_{ST(2,0)}}\right]^2  .
\end{equation}

We do not have a direct experimental estimate for the quantity $\Gamma_{23}^{13}$.  However, we
can estimate it using the parabolic well approximation, where the eigenstates of each dot are those of the 2-dimensional harmonic oscillator~\cite{Davies}. For a dot size of approximately 100~nm and an interdot distance of approximately 200~nm~\cite{Nakul2008}, we obtain $\Gamma_{23}^{13} \approx 0.15$~meV when there is maximum overlap between the first excited eigenstate (`2') of the left dot with the ground eigenstate (`3') of the right dot. This corresponds to the axis of the wavefunction of the first excited eigenstate being aligned with the axis of the double dot. A more realistic estimate where the angle between the wavefunction axis and the axis of the double dot is 45 degrees, gives
\begin{equation}\tag{S28}
\Gamma_{23}^{13} \approx 0.073 \text{ meV}.
\end{equation}
Given that $E_{ST(2,0)} \approx 174 ~\mu$eV~\cite{Christie2010}, the singlet-triplet mixing probability is 
\begin{equation}\tag{S29}
p_m \approx 0.044.
\end{equation}

The cotunneling rate $\Gamma_C$ from the left lead into the right dot and out to the right lead, can be estimated from point F of the transport data (Fig.~S2 and Fig.~1(b) of the main text).
This cotunneling has contributions from both the singlet and triplet channels. The singlet-like channels, $S,T(2,0)\rightarrow S^*(2,1)$, are suppressed by energy dependent tunneling, as is one of the triplet-like channels, $T(2,0)\rightarrow T^*(2,1)$. The other triplet-like channel, $S(2,0)\rightarrow T^*(2,1)$ is suppressed by the slow spin-flip process.
In principle, we cannot separate these contributions at point F.
However, we can consider a worst case scenario where the cotunneling current at point F is dominated by the channels without spin-flip, $S(2,0)\rightarrow S^*(2,1)$ and $T(2,0)\rightarrow T^*(2,1)$, with very little energy dependent tunneling.
In this case, the cotunneling rate through the triplet-like spin-flip channel, $S(2,0)\rightarrow T^*(2,1)$, is given by Eq.~(S26), where $e\Gamma_C$ is the total current measured at point F.  In this way, we obtain
\begin{align}
\Gamma_C\approx 5.0\times10^{5} \text{ s}^{-1},\tag{S30}\\
\Gamma_{S,\text{ unload}}  \approx 2.2\times10^4\text{ s}^{-1}.\tag{S31}
\end{align}

\begin{figure}
\includegraphics[scale=0.25]{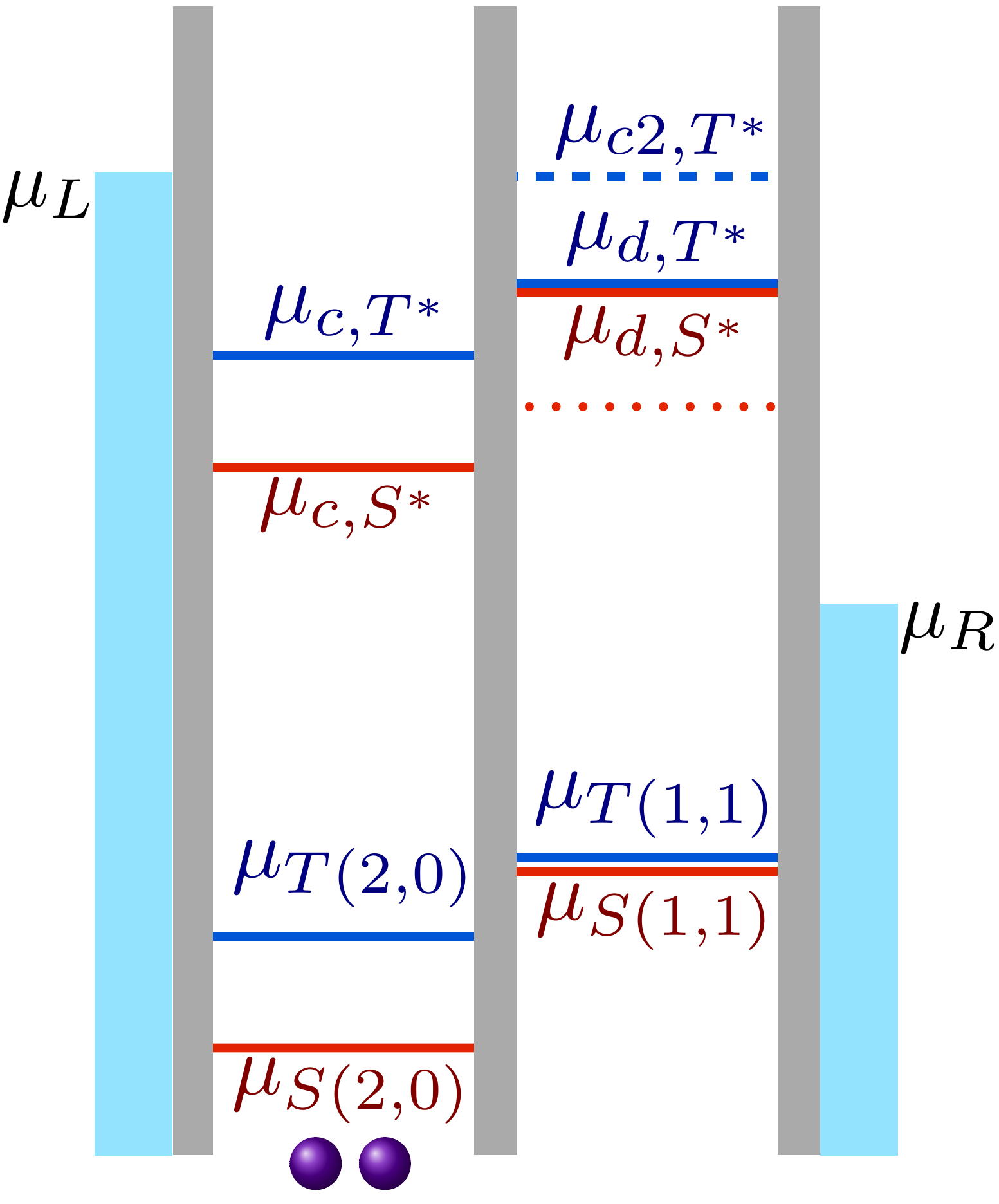}
\caption{\label{fig:F}
Electrochemical potentials at point F of the transport data shown in Fig.~1 of the main text. The dominant current contribution is from cotunneling from the left lead into the right dot and out to the right lead.
 } 
\end{figure}

\subsection{Conclusions}
It is clear from the preceding analysis that the loading and unloading rates satisfy the conditions for Eq.~(S21) to be valid, i.e. $\Gamma_{T\text{ unload}} \gg \Gamma_{T\text{ load}} \gg \Gamma_{S\text{ load/unload}}$. Since the ratio $\Gamma_{S\text{ load}}/\Gamma_{S\text{ unload}} \approx 1$, the current along the tail is of the same order of magnitude as the current in the singlet-like triangle away from the resonant left edge.

Below the tail, cotunneling is energetically unfavorable, while above the tail, cotunneling is suppressed through energy dependent tunneling. $\Gamma_{S\text{ unload}}$ is therefore much smaller on both sides of the tail. Loading of the singlet-like channel, which is strongly energy dependent, increases below the tail, further suppressing current. However, suppression of the singlet-like channel loading rate above the tail could mitigate the decrease in the unloading rate, so that the ratio $\Gamma_{S\text{ load}} / \Gamma_{S\text{ unload}}$ remains small, allowing some current flow. That there is small current above the tail is observed in transport data at larger biases~\cite{Christie2010}, where the loading of the singlet-like channel can be more strongly suppressed as it moves lower in the bias window. 

Therefore, although spin-flip cotunneling involves two mechanisms of lower probability, namely cotunneling and triplet-singlet mixing, the dampening effect on the tunnel rate is mitigated by the small loading probability into the singlet-like channel. The effect of spin-flip cotunneling is to open a thin window in the triplet-like bias triangle to allow triplet-like channel current.


\section{Proposed Experiment to Verify Singlet-Triplet Mixing Mechanism} 

In this section, we propose an experiment to verify the singlet-triplet mixing mechanism described in the main text.  We note that in the spin-flip process described in the text, the loading of the third electron into the right dot is through cotunneling from the left lead. Since it is the singlet-triplet mixing mechanism that we are interested in, this loading can also be from the tunneling of an electron from the right lead, as described below.

\begin{figure}
\includegraphics[scale=0.2]{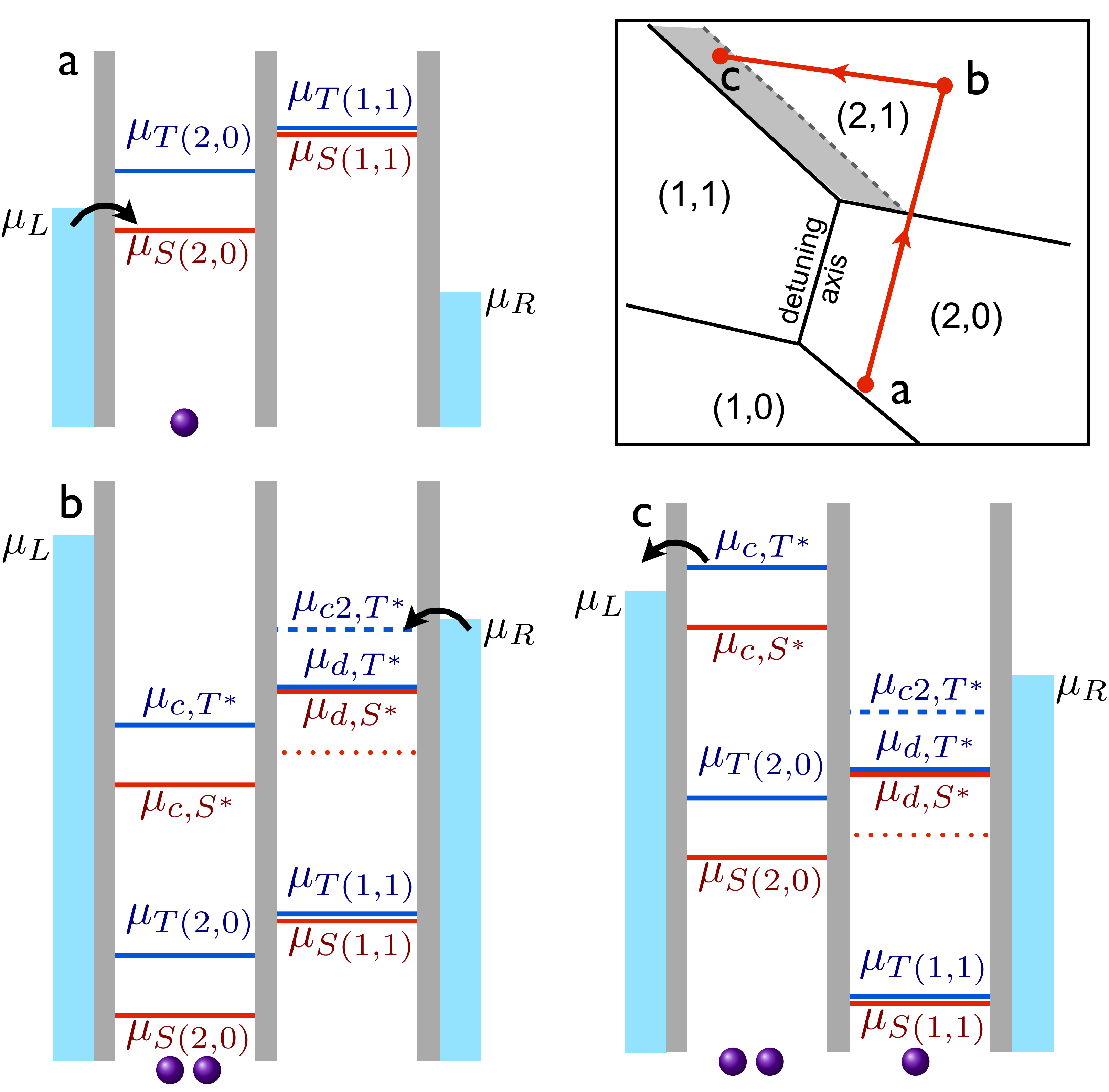}
\caption{\label{fig:expt}
Proposed experimental verification of the singlet-triplet spin-flip process. The boxed panel shows a schematic of the honeycomb charge stability diagram~\cite{vanderWiel}, showing the regions in which the ground (singlet and singlet-like) states of the various charge configurations are stable. Points a, b and c are the positions to which the double dot should be pulsed. The dotted line indicates the boundary between the (1,1) and triplet-like (2,1) regions, i.e. within the shaded region, a (1,1) state is more stable than the triplet-like (2,1) state. Panels a, b and c show the electrochemical potentials, with the circles at the bottom of each diagram indicating the number of electrons at the start of each pulse. Black arrows indicate the tunneling of an electron, which can be detected via charge sensing using an adjacent quantum point contact. Detection of a tunneling out process at `c' would indicate a singlet-triplet spin-flip.}
\end{figure}

Using pulsed gate techniques such as those described in Ref.~\onlinecite{EnDep2}, the double dot could be first loaded into the (2,0) singlet state (a), then pulsed such that the $\mu_{c2,T^*}$ level of the right dot is below the Fermi level of the right lead (b), allowing for the loading of an electron into the right dot from the right lead. Loading of the $\mu_{d,S^*}$ level will not cause singlet-triplet mixing but this is suppressed because of energy dependent tunneling. Next, a gate pulse can then be applied such that the triplet-like level $\mu_{c,T^*}$ is above the Fermi level of the left lead (c). A singlet-triplet spin-flip will lead to a tunneling out of an electron via this level. Tunneling events can be detected by charge sensing via an adjacent quantum point contact~\cite{EnDep2}. These three pulse stages, a to c, are shown schematically in Fig.~\ref{fig:expt}. 


\section{Transport through the Singlet-like channel}

\begin{figure}
\includegraphics[scale=0.17]{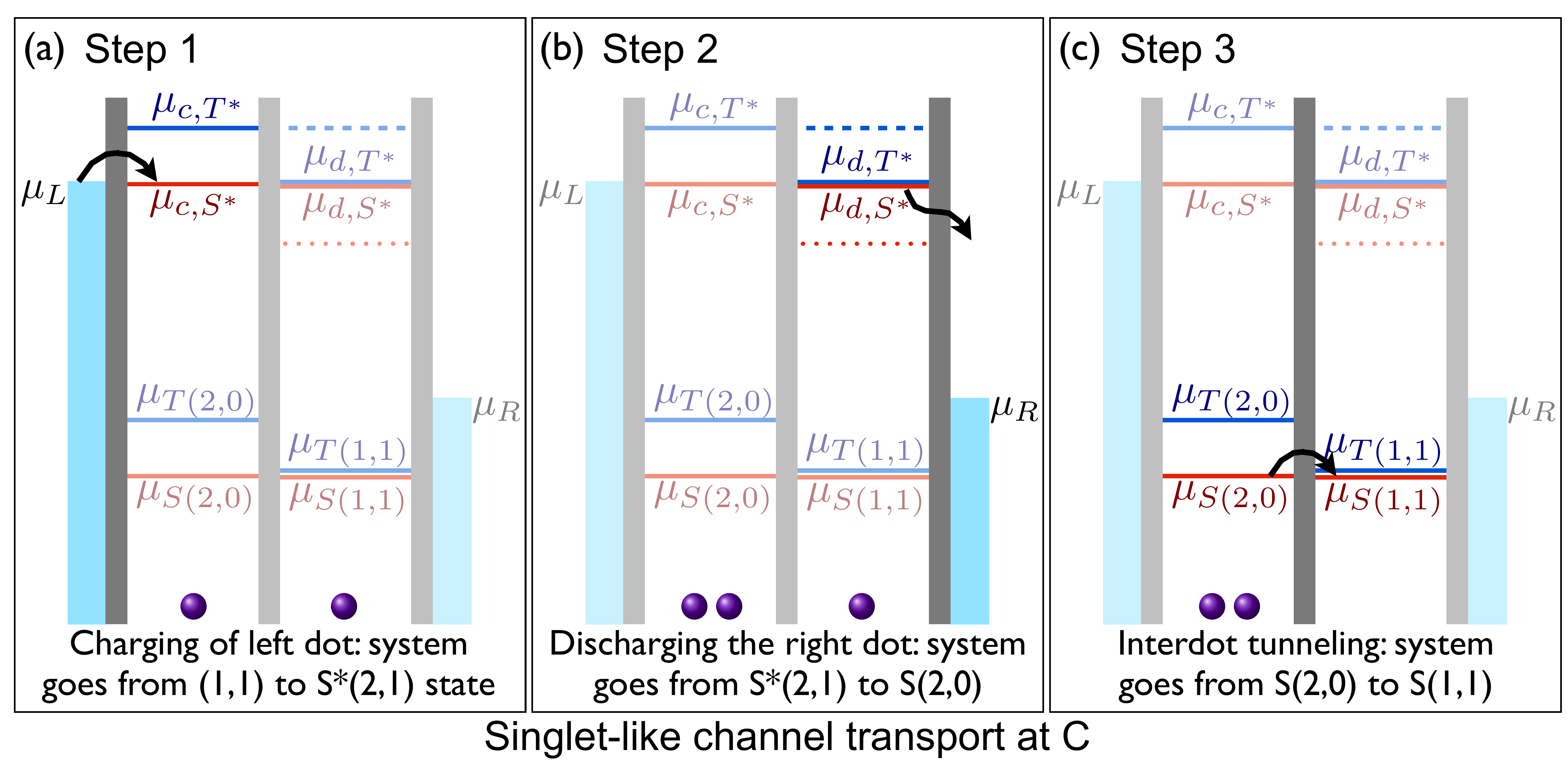}
\caption{\label{fig:levels2} Diagrams of singlet-like channel transport at point C. Purple dots at the bottom of each diagram indicate the electron number at the beginning of each step in the transport cycle. (a) Starting from a (1,1) state, charging of the left dot from the $\mu_{c,S^*}$ level brings the system to the $S^*(2,1)$ state.
(b) From the $S^*(2,1)$ state, the left dot ends up in the $S(2,0)$ state as the right dot discharges from the $\mu_{d,S^*}$ level. (c) Inter-dot tunneling from (2,0) to (1,1) completes the cycle. } 
\end{figure}

The strong line on the left edge of the singlet-like triangle in Fig.~1 of the main text can be explained by Fig.~\ref{fig:levels2}. At C, the $\mu_{c,S^*}$ level of the left dot is resonant with the Fermi level of the left lead, $\mu_{L}$, as shown in Fig.~\ref{fig:levels2}(a).  Also, $\mu_{S(2,0)}$ and $\mu_{S(1,1)}$ are aligned with each other as shown in Fig.~\ref{fig:levels2}(c). These two resonance conditions lead to a sharp current peak at C. Going away from C along the left edge of the triangle (Fig.~1, main text), the right dot falls in energy but the left dot maintains its resonance with the left lead, giving rise to a bright line along the left edge. In the direction where both dot energies fall in tandem, loading of the left dot into $\mu_{c,S^*}$ from the left lead is suppressed due to the exponential energy dependence in tunneling~\cite{Christie2010}.

\section{Spin-flip Cotunneling in Transport Data at Forward Bias} 

Here we discuss the interpretation of transport data taken at forward bias of $V_{\text{SD}} = + 0.526 \text{mV}$, originally reported in Ref.~\onlinecite{Christie2010} and shown in Fig.~\ref{fig:Spin-Relax} and show that certain features of the data arise because of spin-flip cotunneling.

\begin{figure*}[!h]
\includegraphics[scale=0.24]{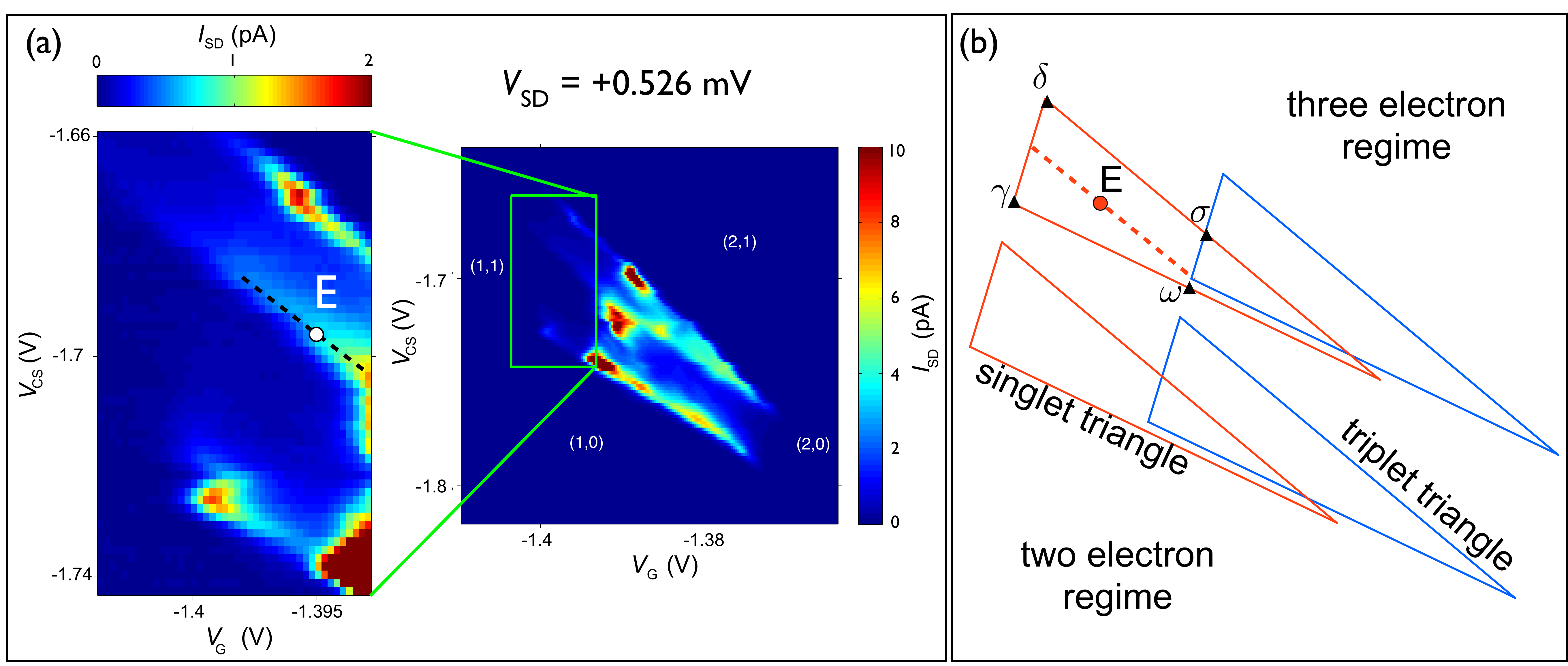}
\caption{\label{fig:FwdBiasData} Transport data and schematic of bias triangles at forward bias. (a) Transport current $I_{\text{SD}}$ as a function of two gate voltages, $V_{\text{G}}$ (V) and $V_{\text{CS}}$ (V), taken at a forward bias source-drain voltage of $V_{\text{SD}} = + 0.526 \text{mV}$. These data were reported originally in Ref.~\onlinecite{Christie2010}. The feature shown by the dotted line is explained by the spin-flip cotunneling process. 
(b) Schematic of the bias triangles of transport data taken at the forward bias. Bias triangles corresponding to transport through ground (excited) states are shown in the red (blue) on the left (right). In the three electron regime (upper triangles), the dotted line lies in the region $\gamma \delta \sigma \omega$, where the triplet $T(1,1)$ state is spin blockaded. Spin relaxation to the $S(1,1)$ state lifts the spin blockade and allows transport to resume. This is possible within the region bounded by $\gamma \delta \sigma \omega$ and should give rise to a uniform current within the region. However, the presence of the line of high current (dotted) within $\gamma \delta \sigma \omega$ suggests a preferentially high spin relaxation rate which is explained by the spin-flip cotunneling process shown below in Fig.~\ref{fig:Spin-Relax}. }
\end{figure*}

\begin{figure*}
\includegraphics[scale=0.2]{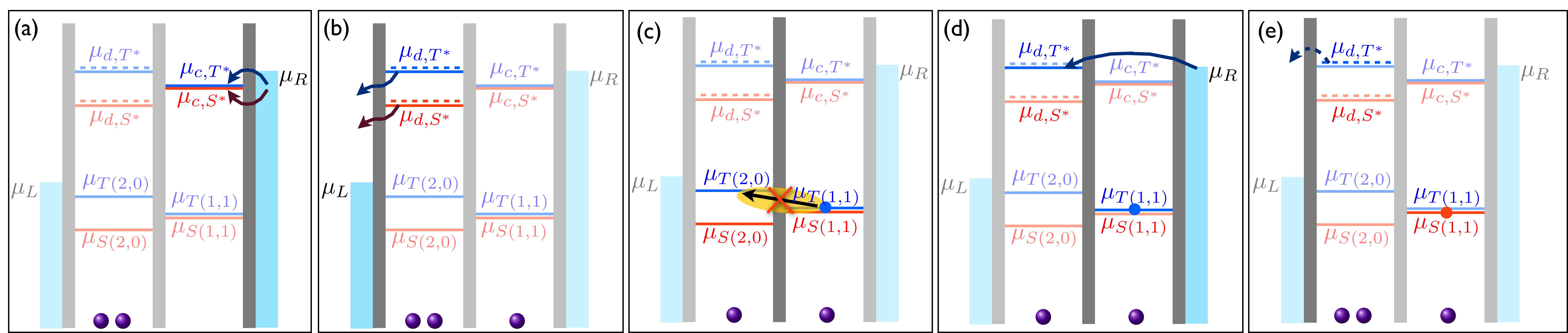}
\caption{\label{fig:Spin-Relax} Processes involved in the lifting of the triplet $T(1,1)$ spin blockade through cotunneling induced spin flip at point E.
In the three-electron regime (upper triangles) of Fig.~\ref{fig:FwdBiasData}(b), transport is spin blockaded within the region bounded by $\gamma \delta \sigma \omega$. Purple dots at the bottom of each diagram indicate the electron number at the beginning of each step in the transport cycle.
(a) Starting from any one of the two (2,0) states, the $S^*(2,1)$ or $T^*(2,1)$ state is loaded from the right lead onto the left dot at the $\mu_{c,S^*}$ or $\mu_{c,T^*}$ chemical potential respectively. 
(b) The left dot may discharge from the lower (solid line) of the $\mu_{d,S^*}$ or $\mu_{d,T^*}$ chemical potential, resulting in a $T(1,1)$ triplet. (Discharge from any one of the two dashed levels is also possible, but it results in a singlet $S(1,1)$ that is not spin blockaded.)
(c) The $T(1,1)$ triplet state is spin blockaded as the $\mu_{T(1,1)}$ chemical potential lies below the $\mu_{T(2,0)}$ level.
(d) Along the dashed line of Fig.~\ref{fig:FwdBiasData}, the lower (solid line) of the two $\mu_{d,T^*}$ chemical potentials is lined up with the Fermi level of the right lead, allowing the loading of the excited (2,1) state through cotunneling from the right lead. 
(e) Subsequent discharge of a left dot electron could be from either one of the $\mu_{d,T^*}$ levels, with discharge from the upper (dashed line) of the $\mu_{d,T^*}$ levels resulting in the singlet $S(1,1)$ state which can then complete the transport cycle by going into the $S(2,0)$ state through inter-dot tunneling. Transport thus resumes until the next triplet (1,1) spin blockade. The key is that from either the singlet-like or triplet-like (2,1) states, discharge from the left dot at the chemical potentials indicated by the dashed or solid lines results in a singlet or triplet (1,1) state respectively. The former allows transport to resume, whereas the latter results in a spin blockade. 
The dotted line of Fig.~\ref{fig:FwdBiasData} lies in the direction where the left dot energy is fixed, leading to the line of high current.
}
\end{figure*}

In the three-electron regime, transport at forward bias cycles from $(1,1)\rightarrow(2,0)\rightarrow(2,1)$. The bias triangles for forward bias are shown in Fig.~\ref{fig:FwdBiasData}(b). In this regime, transport is spin blockaded within the region bounded by $\gamma \delta \sigma \omega$ of Fig.~\ref{fig:FwdBiasData}(b), due to the fact that $\mu_{T(1,1)}$ is below the $\mu_{T(2,0)}$ level as shown in Fig.~\ref{fig:Spin-Relax}(c). Any mechanism that allows spin relaxation to the $S(1,1)$ state lifts the spin blockade, and typical relaxation mechanisms are expected to give rise to uniform current over the entire blockade region $\gamma \delta \sigma \omega$. However, we observe a line of high current, shown by the dotted line in Fig.~\ref{fig:FwdBiasData}, which indicates that there is a preferentially high relaxation rate along this feature. We explain this line of strong current by a spin-flip cotunneling process.

Point E (Fig.~\ref{fig:FwdBiasData}) lies along the line of high current in the spin blockade region. At this point, the lower (solid line) of the two $\mu_{d,T^*}$ levels of the left dot are lined up with the right lead. This allows cotunneling of an electron from the right lead into the left dot to form the triplet-like $T^{*}(2,1)$ state. (Fig.~\ref{fig:Spin-Relax}(d)). Discharging of the left dot from the higher (dashed line) of the $\mu_{d,T^*}$ chemical potential results in the singlet $S(1,1)$ state, which is not blockaded. The double dot is then able to complete the transport cycle until the next time the double dot gets spin blockaded. 

It is because the dotted line in Fig.~\ref{fig:FwdBiasData} lies in the right dot energy axis that the spin blockade is lifted at a much faster rate. This is due to the fact that the left dot energy is aligned with the Fermi level of the right lead, allowing cotunneling to occur at a fast rate. As a result, spin relaxation happens at a much higher rate along this line.


\end{document}